\documentclass[twocolumn]{aa}
\usepackage{txfonts, graphicx}
\usepackage[round]{natbib}
\newcommand{\te}{\ensuremath{T_{\mathrm{eff}}}}
\newcommand{\kms}{km s$^{-1}$}
\newcommand{\vsi}{\ensuremath{v \sin i}}
\newcommand{\vr}{\ensuremath{v_R}}
\newcommand{\bz}{\ensuremath{\langle B_z \rangle}}

\begin{document}
\titlerunning{Si {\sc ii} and Si {\sc iii} abundances in B-type Stars}
\title{Abundances determined using  Si~{\sc ii} and Si~{\sc iii}  in B-type stars: evidence for stratification\thanks{Based in part on observations made with the European Southern Observatory (ESO) telescopes under the ESO programmes 067.D-0579(A), 072.D-0410(A), 076.B-0055(A), 076.D-0169(A), 078.D-0080(A), and 086.D-0449(A), obtained from the ESO/ST-ECF Science Archive Facility. It is also based in part on observations carried out at the Canada-France-Hawaii Telescope (CFHT) which is operated by the National Research Council of Canada, the Institut National des Science de l'Universe of the Centre National de la Recherche Scientifique of France and the University of Hawaii. It is based in part on observations available on the ELODIE archive.}}
\author{J. D. Bailey\inst{1}
\and J. D. Landstreet\inst{1,2}}
\institute{Department of Physics \& Astronomy, The University of Western Ontario, London, Ontario, N6A 3K7, Canada
\and
Armagh Observatory, College Hill, Armagh, Northern Ireland BT61 9DG}
\date{Received October 30, 2012~/~Accepted January 3, 2013}

\abstract {It is becoming clear that determination of the abundance of Si using lines of Si~{\sc ii} and Si~{\sc iii} can lead to quite discordant results in mid to late B-type stars. The difference between the Si abundances derived from the two ion states can exceed one dex in some cases.}
{We have carried out a study intended to clarify which kinds of B stars exhibit this discrepancy, to try to identify regularities in the phenomenon, and to explore possible explanations such as abundance stratification by comparing models to observed spectra.}
{We used spectra from the ESPaDOnS spectropolarimeter and FEROS spectrograph, supplemented with spectra from the ESO and ELODIE archives, of magnetic Bp, HgMn, and normal B-type stars ranging in effective temperature from about 10500 to 15000~K.  Using these spectra, we derived abundances using the spectrum synthesis program \textsc{zeeman} which can take into account the influence of magnetic fields.  For each star, accurate abundances of Si~{\sc ii}, Si~{\sc iii}, Ti, Cr, and Fe were derived from two separate $\sim$100~\AA\ windows.  Si~{\sc ii} abundances were deduced from multiple lines, and Si~{\sc iii} abundances were found using $\lambda\lambda$~4552, 4567, and 4574.}
{All magnetic Bp stars in our sample show a discordance between the derived abundances of the first and second ions of silicon, with the latter being between 0.6 - 1.7~dex higher.  The same behaviour is observed in the non-magnetic stars but to a much smaller extent: Si~{\sc iii} is enhanced by between 0.3 - 0.8~dex compared to Si~{\sc ii}.  We do not detect the discrepancy in three stars, HD~22136 (normal), HD~57608 (HgMn) and HD~27295 (HgMn); these are the only stars in our sample for which the microturbulence parameter is significantly different from zero, and which therefore probably have convection occurring in their atmospheres.}
{We find that vertical stratification of silicon in the atmospheres of B-type stars may provide an explanation of this phenomenon, but our detailed stratification models do not completely explain the discrepancies, which may, in part, be due to non-LTE effects.}

\maketitle

\section{Introduction}
The magnetic peculiar A and B-type stars (Ap/Bp) are characterized by large overabundances of Fe-peak and rare-earth elements.  They possess strong, non-axisymmetric magnetic fields. The observed line-of-sight magnetic field component, \bz, is typically of the order of 1~kG or more in strength, and usually varies as the star rotates.  The abundance ratios of the chemical elements are often variable over the stellar surface (``patchy''), leading to line strength and shape variations with stellar rotation.  This situation is possible because the magnetic field inhibits any convective or turbulent mixing.  

In addition to this, \citet{Wade2001} argued that most magnetic stars exhibit vertical abundance variations (chemical stratification), an idea first explored by \citet{Babel1992}. The phenomenon was further studied by \citet{Ryabchikova2001}, who compared the chemical abundances derived from the singly- and doubly-ionised forms  of Pr and Nd in magnetic and non-magnetic A-type stars. In pulsating (``roAp'') magnetic stars, abundances derived from the second ions were found to be 1-1.7~dex larger than abundances from the first ions. In contrast, they found no significant discrepancies between the abundances derived from the first and second ionization states of Pr and Nd in the non-magnetic and non-pulsating stars. Lines of Pr~{\sc iii} and Nd~{\sc iii} would normally be formed deeper in the atmospheres of A stars than the lines of the singly ionised rare earths, because of the higher local temperature deep in the atmosphere.  However, as \citet{Ryabchikova2001} pointed out, since electron number density decreases outward, the second ionization state can also dominate in the very uppermost atmospheric layers. Because of this effect, they were able to understand the discrepancy between ion states as being due to large rare earth overabundances concentrated in a thin layer high in the atmosphere. 

A related discrepancy has been observed for abundances of Si derived from two ion states. \citet{Semenko2008} studied the magnetic Bp star HD~45583, which has effective temperature $\te = 13000$~K, $\log g = 4.0$ and a strong magnetic field, with \bz\ varying from about $-3500$ to $+4000$~G.  Different abundances must be assumed for Si~{\sc ii} and Si~{\sc iii} to adequately fit the theoretical to the observed line profiles. They interpret this discrepancy as being due to vertical stratification of silicon in the atmosphere, but point out that increasing \te\ from the assumed value also brings the abundances derived from lines of Si~{\sc ii} and Si~{\sc iii} into better agreement.  \citet{Bailey2012} found a similar discordance in the derived abundances of the first and second ionization states of Si in HD~133880, another strongly magnetic Bp star (\bz\ varies from about $-4000$ to $+2000$~G) having $\te = 13000$~K and $\log g = 4.34$. They find that an enhancement of about 1~dex above the Si~{\sc ii} abundance is necessary to adequately fit the Si~{\sc iii} lines.  Since non-LTE effects in the silicon lines are expected to be small below a \te\ of about 15000~K, they suggest that the most probable cause is strong stratification of Si in the atmosphere. 

The study of vertical abundance stratification in magnetic Ap stars has considerably expanded in the past few years with work led by Kochukhov and Ryabchikova \citep[e.g.][]{Kochukhov2006b,Kochukhov2009,Shulyak2009,Pandey2011}. This group has developed refined methods of recovering the vertical distribution of several elements in a stellar atmosphere by simultaneous fits to a number of spectral lines of different strengths, excitation potentials, and ionisation states. They have studied several cool magnetic Ap stars in detail, and mapped the vertical distribution of a number of elements in each, in several cases including Si. Typically, it is found that Si has high abundance deep in the atmosphere, and an abundance that declines strongly towards low optical depths, sometimes by as much as 3~dex overall. 

Up to now, the study of discrepancies among different ion states, and of stratification as a probable explanation, has mainly been confined to cool magnetic Ap stars. It is time for a first survey of higher effective temperatures. In this paper, we report an empirical study of abundances derived from lines of Si~{\sc ii} and Si~{\sc iii} in various classes of B-type stars including magnetic Bp, HgMn, and normal B stars.  Stars were selected from the ESO, CFHT, and ELODIE archives (as well as from our own data from previous CFHT and ESO programmes). The stars studied have \te\ values that range from about 10500 to 15000~K, and have low enough projected rotational velocity \vsi\ (less than about 100~\kms) that lines of Si~{\sc iii} can be unambiguously detected.  The following section discusses the spectroscopic observations used.  Sect.~3 outlines the modelling technique and the spectrum synthesis program used.  Sect.~4 presents the abundances we derive from lines of Si~{\sc ii} and Si~{\sc iii}.  Sect.~5 investigates the possible effects of non-LTE in deriving abundances of Si~{\sc ii} and Si~{\sc iii} with increasing \te\ and we test whether the observed discrepancies can be accounted for using stratified abundance models. Sect.~6 summarises and discusses the work we present.

\section{Stellar sample}

\begin{table}
\centering
\caption{Stars analyzed in this study.  Listed are the star designations, instrument used, spectral resolution, spectral range and the S/N.}
\begin{tabular}{lrrrr}
\hline\hline
Star & Instrument & R & $\lambda$ (\AA) & S/N \\
\hline
\multicolumn{5}{c}{normal stars}\\
HD 22136 & ESPaDOnS & 65000 & 3690-10481 & 504 \\
HD 160762 & ESPaDOnS & 65000 & 3690-10481 & 600 \\
HD 162586 & UVES & 110000 & 3070-10398 & 655 \\
HD 170054 & ESPaDOnS & 65000 & 3690-10481 & 283\\
HD 179761 & ELODIE & 42000 & 4000-6800 & 125\\
HD 195810 & ESPaDOnS & 65000 & 3690-10481 & 774\\ 
HD 222173 & ELODIE & 42000 & 4000-6800 & 159 \\
\hline
\multicolumn{5}{c}{HgMn stars}\\
HD 27295  & UVES & 110000 & 3750-6800 & 218 \\
HD 57608  & FEROS & 48000 & 3528-9217 & 142\\
HD 78316  & UVES & 110000 & 3750-6800 & 265\\
HD 175640  & UVES & 110000 & 3750-5800 & 439 \\
HD 178065  & UVES & 110000 & 3750-5800 & 461\\
HD 186122  & ESPaDOnS & 65000 & 3690-10481 & 662\\
HD 193452 & CF4 & 120000 & 4630-5478 & 556\\
\hline
\multicolumn{5}{c}{magnetic Bp stars}\\
HD 10840 & UVES & 110000 & 3070-10398 & 824 \\  
HD 45583 & FEROS & 48000 & 3528-9217 & 231\\   
HD 47116 & UVES & 110000 & 3070-10398 & 527 \\ 
HD 49333 & UVES & 110000 & 3070-10398 & 791 \\   
HD 61045 & ESPaDOnS & 65000 & 3690-10481 & 298 \\  
HD 74168 & UVES & 110000 & 3070-10398 & 334\\   
HD 74535 & FEROS & 48000 & 3528-9217 & 310 \\  
HD 133880 & ESPaDOnS & 65000 & 3690-10481 & 365\\  
HD 137509 & UVES & 110000 & 3690-10481 & 430\\ 
HD 147010 & ESPaDOnS & 65000 & 3690-10481 & 587\\  
HD 199728 & UVES & 110000 & 3070-10398 & 593\\ 
HD 223640 & UVES & 110000 & 3070-10398 & 560\\
HD 304842 & FEROS & 48000 & 3528-9217 & 262 \\
HD~318107 & ESPaDOnS & 65000 & 3690-10481 & 320\\ 
BD+00 1659 & ESPaDOnS & 65000 & 3690-10481 & 475 \\
BD-19 5044L & ESPaDOnS & 65000 & 3690-10481 & 358 \\
BD+49 3789 & ESPaDOnS & 65000 & 3690-10481 & 191\\ 
\hline
\label{observations}
\end{tabular}
\end{table}

We wish to study the extent to which the abundance of Si as derived from Si~{\sc ii} and Si~{\sc iii} is concordant or discordant. With the tools readily available to us, this can only be done in a limited range of \te\ values or spectral range. Our low-temperature limit is set by the fact that no Si~{\sc iii} lines are visible in optical spectra for \te\ less than about 11\,000~K. The high-temperature limit is set by the gradual onset of non-LTE effects at about 15\,000~K \citep{Przybilla2011}. This upper limit to our study, which is not well-defined, is imposed by two main  factors. First, the non-LTE codes developed up to now do not incorporate radiative transfer in the presence of a magnetic field. Thus, they cannot be used directly to study magnetic stars which are an important class of objects for this study. Furthermore, as we will show below, the published grid of non-LTE equivalent widths for Si~{\sc ii} and {\sc iii} lines of \citet{Becker1990} do not even approximately reproduce the line ratios in the normal B3 IV star $\iota$~Her = HD 160762, and recent non-LTE abundance studies of early B stars appear to systematically reject use of the Si~{\sc ii} lines. It is not at all clear that current non-LTE codes are able to correctly compute line ratios for Si~{\sc ii} lines. Therefore we try to limit our study to effective temperatures low enough that strong non-LTE effects are not expected; this limiting temperature is conventionally set at $\te = 15000$~K \citep[e.g.][]{Nieva2012}. We have included one hotter star, HD~160762, in order to address non-LTE questions later in the paper. 

The spectra for this study were obtained from a variety of sources.  A large portion of the stars studied were acquired from previous observing runs using the ESPaDOnS spectropolarimeter (with resolving power $R = 65000$) at the Canada-France-Hawaii Telescope (CFHT), and the FEROS spectrograph ($R = 48000$) at the European Southern Observatory's (ESO) La Silla Observatory.  The remaining spectra used were acquired from the ESO, CFHT and ELODIE (Observatoire de Haute Provence) archives.  We required that each individual spectrum have a signal-to-noise ratio (S/N) greater than about 100 and spectral resolution no less than about 40~000.  As discussed above, we selected stars of \te\ between about 10500 and 15000~K, requiring also that they have sufficiently slow rotation (\vsi\ less than about 100~\kms) that at least some individual lines are not strongly blended.  

In the effective temperature range of interest for this study, there are (at least) three major classes of main sequence stars. In addition to normal late B stars (which generally do not have detectable magnetic fields, and have approximately solar surface abundances), we also find magnetic Bp stars, which usually show quite marked abundance differences compared to the Sun,  and the non-magnetic HgMn stars, which are typically found in fairly close binary systems and show quite non-solar over-abundances of several low-abundance elements such as P, Mn, Ga, and Hg. We have included several examples of each of these classes in our sample. The sample of stars utilized in this study is summarised in Table~\ref{observations}.

The fundamental parameters \te\ and $\log g$ were derived for each star using Geneva and $uvby\beta$ photometry.  For the Geneva photometry, we use the {\sc fortran} program described by \citet{Kunzli1997}.  For stars with Str\"{o}mgren $uvby\beta$ photometry, the  {\sc fortran} program ``UVBYBETANEW'' of \citet{Napiwotzki1993}. Our version of this program corrects the \te\ of the magnetic Ap/Bp stars to a suitable Ap temperature scale  \citep[see][]{Landstreet2007}.  For stars for which both sets of photometry were available, the average value was taken. We were able to compare our \te\ values with those of \citet{Netopil2008} in a few cases; our values agree with theirs except for HD~133880, for which \te\ is somewhat more uncertain, as discussed by \citet{Bailey2012}.  Finally, we have estimated the uncertainty in \te\  to be about 500~K, following the discussion of \citet{Landstreet2007}. The uncertainty in $\log g$ is estimated from the level of agreement between values derived from Geneva and Str\"{o}mgren photometry to be about 0.2~dex.  The physical parameters for HD~160762 were taken from \citet{Nieva2012}, and have uncertainties that are about half as large as those of the other normal stars.  Table~\ref{parameters} summarises all the physical parameters of the stars modelled in this study, including parameters discussed later in this paper.

\section{Modelling spectra}
\subsection{Modelling with the spectrum synthesis code {\sc zeeman}}
\begin{table*}
\centering
\caption{Physical parameters for the stars in this study.  We adopt a uniform uncertainty of $\pm$500~K in \te\ and $\pm$0.2 in $\log \rm g$. }
\begin{tabular}{lrrrrrrrrr}
\hline\hline
Star & \te\ (K) & $\log \rm g$ & \vsi\ (\kms) & $i$ ($^{\circ}$) & $\beta$ ($^{\circ}$) & B$_{d}$ (G) & B$_{q}$ (G) & B$_{oct}$ (G) & $\xi$ (\kms) \\  
\hline
\multicolumn{10}{c}{normal stars}\\
HD 222173 & 11800 & 3.4 & 67 $\pm$ 5 & -- & -- & -- & -- & -- & $<$ 1.4 \\
HD 22136 & 12700 & 4.2 & 15 $\pm$ 2 & -- & -- & -- & -- & -- & 1.1 $\pm$ 0.3 \\
HD 162586 & 12700 & 4.0 & 28 $\pm$ 2 & -- & -- & -- & -- & -- & $<$ 1.6 \\
HD 179761 & 12900 & 3.4 & 14 $\pm$ 2  & -- & -- & -- & -- & -- & $<$ 1.5 \\
HD 195810 & 13700 & 3.7 & 52 $\pm$ 4 & -- & -- & -- & -- & -- & $<$ 1.1\\
HD 170054 & 14500 & 4.3 & 25 $\pm$ 2 & -- & -- & -- & -- & -- & $<$ 1.2 \\
HD 160762 & 17500 & 3.8 &  6  $\pm$ 1 &  -- & -- & -- & -- & -- &  1 $\pm$ 1 \\
\hline
\multicolumn{10}{c}{HgMn stars}\\
HD 193452 & 10600 & 4.1 & 1.4 $\pm$ 1 & -- & -- & -- & -- & -- & $<$ 1.0\\
HD 57608 & 10900 & 3.1 & 5.0 $\pm$ 1 & -- & -- & -- & -- & -- & 1.3 $\pm$ 0.4 \\
HD 27295 & 11800 & 4.2 & 4.9 $\pm$ 0.3 & -- & -- & -- & -- & -- & 1.8 $\pm$ 0.5\\
HD 175640 & 12000 & 4.0 & 1.5 $\pm$ 1 & -- & -- & -- & -- & -- & $<$ 1.7\\
HD 178065 & 12300 & 3.6 & 2.0 $\pm$ 1 & -- & -- & -- & -- & -- & $<$ 1.5\\
HD 186122 & 12900 & 3.7 & 1.0 $\pm$ 0.5 & -- & -- & -- & -- & -- & $<$ 1.4\\
HD 78316 & 13400 & 3.9 & 6.8 $\pm$ 0.5 & -- & -- & -- & -- & -- & $<$ 1.6\\
\hline
\multicolumn{10}{c}{magnetic Bp stars}\\
HD 47116   & 11000 & 4.1 & 30 $\pm$ 2 & -- & -- & 500 & -- & -- & 0 \\
HD 10840   & 11600 & 3.6 & 35 $\pm$ 5 & -- & -- & 500 & -- & -- & 0\\
HD~318107  & 11800 & 4.2 & 7 $\pm$ 1 & 22 & 65 & 25600 & $-12800$ & 900 & 0\\
HD 199728  & 12200 & 3.7 & 62 $\pm$ 6 & -- & -- & 800 & -- & -- & 0\\
HD 223640  & 12300 & 4.4 & 31 $\pm$ 3 & -- & -- & 500 & -- & -- & 0\\
BD+00 1659 & 12500 & 4.0 & 7.0 $\pm$ 1 & -- & -- & 1200 & -- & -- & 0\\
HD 304842  & 12500 & 3.9 & 65 $\pm$ 5 & -- & -- & 100 & -- & -- & 0\\
HD 45583   & 12700 & 4.2 & 70 $\pm$ 6 & -- & -- & 8000 & -- & -- & 0\\
BD-19 5044L & 12800 & 4.5 & 15 $\pm$ 3 & -- & -- & 800 & -- & -- & 0\\
BD+49 3789 & 12900 & 4.2 & 85 $\pm$ 5 & -- & -- & 1700 & -- & -- & 0\\
HD 74168   & 12900 & 4.5 & 63 $\pm$ 4 & -- & -- & 200 & -- & -- & 0 \\
HD 61045   & 13000 & 4.1 & 64 $\pm$ 3 & -- & -- & 1300 & -- & -- & 0\\
HD 147010  & 13000 & 4.4 & 15 $\pm$ 2 & -- & -- & 15000 & -- & -- & 0\\
HD 133880  & 13000 & 4.3 & 103 $\pm$ 10 & 55 & 78 & $-9600$ & $-23000$ & 1900 & 0\\
HD 137509  & 13100 & 4.3 & 20 $\pm$ 2 & 81 & 64 & 3100 & 41900 & $-500$ & 0\\
HD 74535   & 13600 & 4.3 & 45 $\pm$ 4 & -- & -- & 300 & -- & -- & 0\\
HD 49333   & 15100 & 3.9 & 68 $\pm$ 3 & -- & -- & 800 & -- & -- & 0 \\
\hline
\label{parameters}
\end{tabular}
\end{table*}

\begin{figure}
\begin{center}
\includegraphics[angle=-90,width=0.5\textwidth]{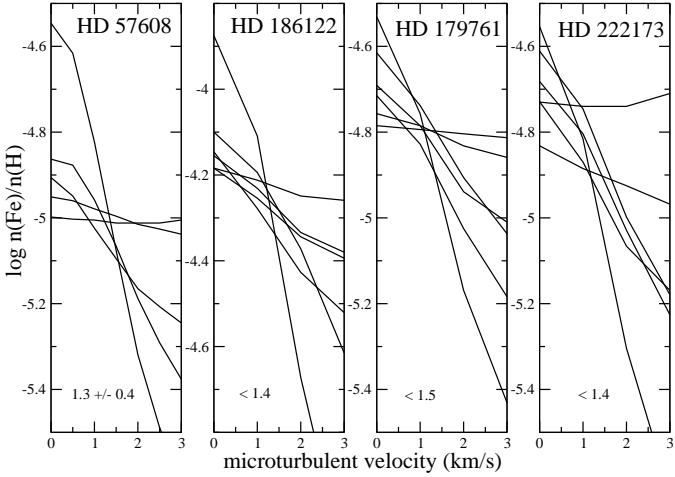}
 \caption{Blackwell diagrams (from left to right) for HD~57608, HD~186122, HD~179761 and HD~222173.  Shown are lines for Fe~{\sc ii} $\lambda\lambda$ 4541, 4555, 4576, 4583, 5018 and sometimes 6120 and/or 6129.  Each panel spans 1.0~dex.} 
\label{xi}
\end{center}
\end{figure}

Modelling and abundance determination was carried out using the {\sc fortran}  spectrum synthesis code {\sc zeeman} \citep[see][]{Landstreet1988, Landstreet1989, Wadeetal2001, Bailey2012}.  {\sc zeeman} accepts as input an assumed magnetic field geometry for stars with magnetic fields (such as the magnetic Ap/Bp stars), but functions as a conventional line synthesis program for modelling of non-magnetic stars \citep[e.g.][]{Landstreet2009}.  The program assumes LTE atmospheric structure and line formation. An appropriate stellar atmosphere model is interpolated in a grid of ATLAS 9 solar abundance models.  For this study, a uniform abundance is assumed over the stellar atmosphere even for stars such as magnetic Bp stars for which this may be only a rough first approximation.  

{\sc zeeman} functions by comparing a synthesized spectrum to an observed one in one or more selected wavelength windows. By iteratively minimizing the mean square difference between the computed and observed spectrum, {\sc zeeman} automatically determines the best fit radial velocity \vr\ and the best value of \vsi\ for the spectral windows modelled. It optimises the fit to all the lines of a single chemical element within the selected window(s), determining the value of the abundance of that element relative the H. This process is repeated for other elements until most of the lines in each spectral window studied are modelled as well as possible with the limitations imposed by the assumptions built into the model. The adopted abundances are obtained from the average of two $\sim$100~\AA\ windows, with the uncertainty estimated from the observed scatter of the computed values in the modelled spectral windows.   

For non-magnetic stars, the microturbulence parameter $\xi$ was determined using the method described by \citet{Landstreet1998}.  The abundance is iteratively deduced for a number of individual lines of Fe for a small grid of assumed $\xi$ values. The resulting abundance values are plotted in a Blackwell diagram for each line as a function of $\xi$. The value of $\xi$ for which the abundance values from the lines studied agree best is accepted as the optimal value, and the corresponding value of Fe/H abundance is adopted. Examination of the dispersion of the abundance values as a function of $\xi$ yields estimates of the uncertainties of both Fe abundance and $\xi$. Several examples of this procedure are shown in Figure~\ref{xi}. 

In most cases, the microturbulence parameter does not seem to be significantly different from zero, but we think that we detect a non-zero value in three stars: the normal star HD~22136, and the two HgMn stars HD~57608 and HD~27295. All three of these stars are near the low temperature limit of our sample, and all three have rather sharp lines, making abundance analysis more precise and perhaps making detection of non-zero $\xi$ easier. Detection of non-zero microturbulence is quite interesting, as it suggests the presence of convective mixing in at least part of the atmosphere \citep[e.g.][]{Landstreet1998,Landstreet2009}. This would in turn suggest that the atmosphere is at least partly mixed, so that a stratified abundance distribution could be destroyed by rapid mixing, if the mixing extends through enough of the atmosphere. 

For magnetic stars, the microturbulence parameter $\xi$ was set to 0, as the magnetic field is thought to suppress convective motions.  

For the magnetic stars  HD~137509, HD~318107, and HD~133880, detailed colinear multipole magnetic field geometries have been reported in the literature that could be used for modelling: \citep[see][respectively]{Koch2006, Bailey2011, Bailey2012}.  For other stars, the line-of-sight magnetic field measurements, \bz,  of \citet{Bagnulo2006,Kochukhov2006,Landstreet2008} and available unpublished data were used to estimate the polar field strength of a simple dipole field model.  In these cases, a dipolar field of about three times the root-mean square of the available \bz\ measurements was used in computing the synthetic spectra.  For stars for which abundances of Fe-peak and rare earth elements clearly indicate a Bp nature, but for which no \bz\ measurements are available, a generic polar field of $\sim$500~G was used.  For all stars for which no previous magnetic modelling was available, we set the inclination of the rotation axis to the line of sight, $i$, and the obliquity of the magnetic field axis to the rotation axis, $\beta$, to zero ($i$=$\beta$=0). 

\subsection{Atomic data}

The atomic data (energy levels, oscillator strengths, damping constants, and Land\'{e} factors) used for this analysis were taken from the Vienna Atomic Line Database (VALD) \citep{vald3, vald2, vald4, vald1}. When Land\'{e} factors were not available from this source, they were computed assuming LS coupling. 

Since it is clear that this study depends sensitively on the accuracy of the atomic line data, particularly oscillator strengths, we have endeavoured to test the VALD $gf$ values extensively. 

One test of the Si~{\sc ii} $gf$ values is to use them for determination of the abundance of silicon in the star Sirius = HD~48915, for which very accurate fundamental and other parameters (e.g. the value of $\xi$) are available, and for which a number of excellent UVES high-resolution spectra are available from the ESO archive \citep{Landstreet2011}. Abundances of Si~{\sc ii} were determined from the lines $\lambda\lambda$~4621, 5041 \& 5056, 5957 \& 5978, and 6347 \& 6371. The derived abundances have a mean of $\log N_{\rm Si}/\log N_{\rm H} = -4.25 \pm 0.07$~dex, where the uncertainty is the dispersion of the four values. The value, about 0.2~dex larger than the solar abundance, is not unexpected in this hot Am star, and the dispersion indicates that the $gf$ values used are closely compatible. 

\begin{table*}
\centering
\caption{Derived abundances from individual lines of Si~{\sc ii} for a selection of stars in our sample.}
\begin{tabular}{lrrrrrrrrrrrrrrr}
\hline\hline 
 & & \multicolumn{4}{c}{HgMn stars} & & \multicolumn{4}{c}{Normal stars} & & \multicolumn{4}{c}{Magnetic stars} \\
\cline{3-6}\cline{8-11}\cline{13-16}
 & & \multicolumn{4}{c}{HD} & & \multicolumn{4}{c}{HD} & & \multicolumn{4}{c}{HD} \\
Line (\AA) & & 178065 & 27295 & 57608 & 78316 & & 222173 & 22136 & 170054 & 162586 & & 74168 & 49333 & 223640 & 47116\\
\cline{1-1}\cline{3-6}\cline{8-11}\cline{13-16}
4621 &   & $-4.67$ & $-4.52$ & $-4.89$ & $-4.34$  & & $-4.66$ & $-4.67$ & $-4.64$ & $-4.67$ & & $-4.00$ & $-3.90$ & $-3.09$ & $-3.21$ \\
5041-55-56 & & $-4.56$ & $-4.58$ & $-4.78$ & $-4.44$  & & $-4.63$ & $-4.64$ & $-4.81$ & $-4.82$ & & $-4.11$ & $-4.15$ & $-3.20$ & $-3.48$ \\
5957-78 &  & --    & $-4.54$ & $-4.66$ & $-4.26$  & & $-4.48$ & $-4.59$ & $-4.70$ & $-4.72$ & & $-3.93$ & $-3.91$ & $-2.98$ & $-3.27$\\
\hline
Dispersion & & --  & 0.03   &  0.12  &  0.09   & &   0.10 &  0.04  &  0.09  &  0.08  &  &  0.09 &  0.14  &  0.11  &  0.14\\
\hline
\label{si2-abundances}
\end{tabular}
\end{table*}

This test has also been carried out on several of the stars in our sample, to see how consistently the VALD data give the same abundances with different spectral lines and line groups. To obtain a realistic idea of the uncertainties in deriving abundances for Si~{\sc ii}, we select a sample of magnetic, normal, and HgMn stars that are well spread in \te\ (within the limitations of the data at hand).  For these stars we derive abundances from $\lambda$4621 individually, $\lambda\lambda$5041, 5055 and 5056 simultaneously, and $\lambda\lambda$5957 and 5978 simultaneously.  Table~\ref{si2-abundances} presents our results, line by line, and with the computed dispersion of the three abundance values for each star.  In general, the abundances we derive from the various lines of Si~{\sc ii} show dispersion which is only slightly greater than that found for Sirius.  The magnetic stars show the greatest tendency for the dispersion of  abundances derived from different lines of Si~{\sc ii} to be larger than we found for Sirius; however, this result is expected, as the magnetic stars typically have horizontally non-uniform abundances and may also have vertically stratified abundance distributions \citep{Wade2001}.  Nevertheless, we note that in general assuming a uniform abundance of Si fits all lines of Si~{\sc ii} in all stars analyzed within a fitting uncertainty of about $\pm$0.15~dex or better.  

A further test of the atomic data is to compare the VALD $gf$ values with the data in the comprehensive NIST compilation of atomic data for Si by \citet{Kelleher2008}. In this data set, the recommended atomic data are selected on a line-by-line basis from available experiments and computations, without consideration of internal homogeneity. Most of the $\log gf$ values of Si~{\sc ii} needed for our work are estimated by those authors to have uncertainties of around $\pm 0.1$~dex. Overall, for the lines studied by us, there is only a very small difference (NIST $-$ VALD) in the mean of all required $\log gf$ values of about $-0.04$~dex for Si~{\sc ii}, and $+0.1$~dex for Si~{\sc iii}. As we shall see below, this is not a large enough difference to play an important role in measurements of the difference in abundances found in some stars between Si~{\sc ii} and Si~{\sc iii}, although if we used the NIST $gf$ values instead of VALD data, this would tend to decrease the abundance differences found by $\sim 0.15$~dex.

We have repeated the Si abundance determinations for Sirius with the NIST Si~{\sc ii} data. The average abundance found is in this way is $-4.20 \pm 0.14$. The change in the mean is thus very small, but the dispersion from line group to line group is doubled with respect to the results found using the VALD data. This kind of difference has been found in earlier comparisons of NIST and other, more homogenized, data \citep[e.g.][]{Sigut1990}, and appears to arise from the NIST preference for data selection for individual lines from single data sources. In particular, the NIST $gf$ value for the Si~{\sc ii} line at 5041~\AA\ is quite discrepant with the VALD value, and when the NIST values are used for synthesis of the spectral region containing the $\lambda\lambda$~5041 and 5056 lines, the best fits are clearly not concordant between the two lines. It is because of this larger dispersion that we prefer to use the VALD atomic data. 

For Si~{\sc iii} we are able to determine the Si abundance using only the lines of multiplet (2) at 4552, 4567, and 4574~\AA.  The $\log gf$ values adopted by VALD, and by us, for these three lines are all about 0.1~dex smaller than those in the NIST compilation. The relative strengths of the lines are closely identical in both datasets. For consistency with our Si~{\sc ii} analyses, we use the VALD values. 

We have also double-checked that our assumed ionisation potentials are very closely the same as those of \citet{Kelleher2008}. In summary, there is no obvious reason to think that the results we find below are due to some major defect in the available spectral line data.

\subsection{Si abundance uncertainties resulting from errors in stellar parameters}

The uncertainty in $\log g$ and especially in \te\ introduces significant uncertainty in the derived abundances, especially of Si~{\sc iii}. This issue has been explored for the different classes of stars in this study by repeating the determination of the best fitting abundance of several stars with values of \te\ or $\log g$ altered by the adopted uncertainties, and comparing the resulting abundance values with those derived with the adopted stellar parameters. 

Over most of the temperature range studied, Si~{\sc ii} is the dominant form of Si throughout much of the atmosphere. Therefore, changing the value of \te\ by 500~K typically results in a fairly small change in the abundance of Si~{\sc ii} which best fits the observed lines. Generally, the change in derived abundance is of the order of 0.05~dex, with the derived abundance decreasing slightly if the adopted \te\ value is increased. Apparently the increased population of the high excitation lower levels of the Si~{\sc ii} lines (at around 10~eV) more than compensates for the effect of the Saha equation in decreasing the fraction of Si in the form of Si~{\sc ii}. The changes to the abundance derived from lines of Si~{\sc ii} for a change in $\log g$ of 0.2~dex is also of order 0.05~dex, with an increase in $\log g$ resulting in a decrease in the abundance deduced from Si~{\sc ii}, as the Saha equation forces more Si from the doubly into the singly ionised state.

In the temperature range studied, the fraction of Si~{\sc iii} increases strongly from a very small fraction at the low temperature end to about half of the total Si near 15000~K. Furthermore, the lower levels of the lines of multiplet (2) are at almost 19~eV. As a result, the abundance of Si derived from lines of Si~{\sc iii} is a much stronger function of \te. With a change of \te\ of $+500$~K, the deduced Si abundance can change by as much as roughly $-0.3$~dex, decreasing somewhat towards the high end of the temperature range studied. The deduced abundance also changes more with a change in $\log g$ of 0.2~dex than is the case for Si~{\sc ii}; the derived abundance can change by an amount of order 0.1~dex.

\section{Measured Si~{\sc ii} and {\sc iii} abundances in the B star sample}
\begin{figure}
\begin{center}
\includegraphics[angle=-90,width=0.5\textwidth]{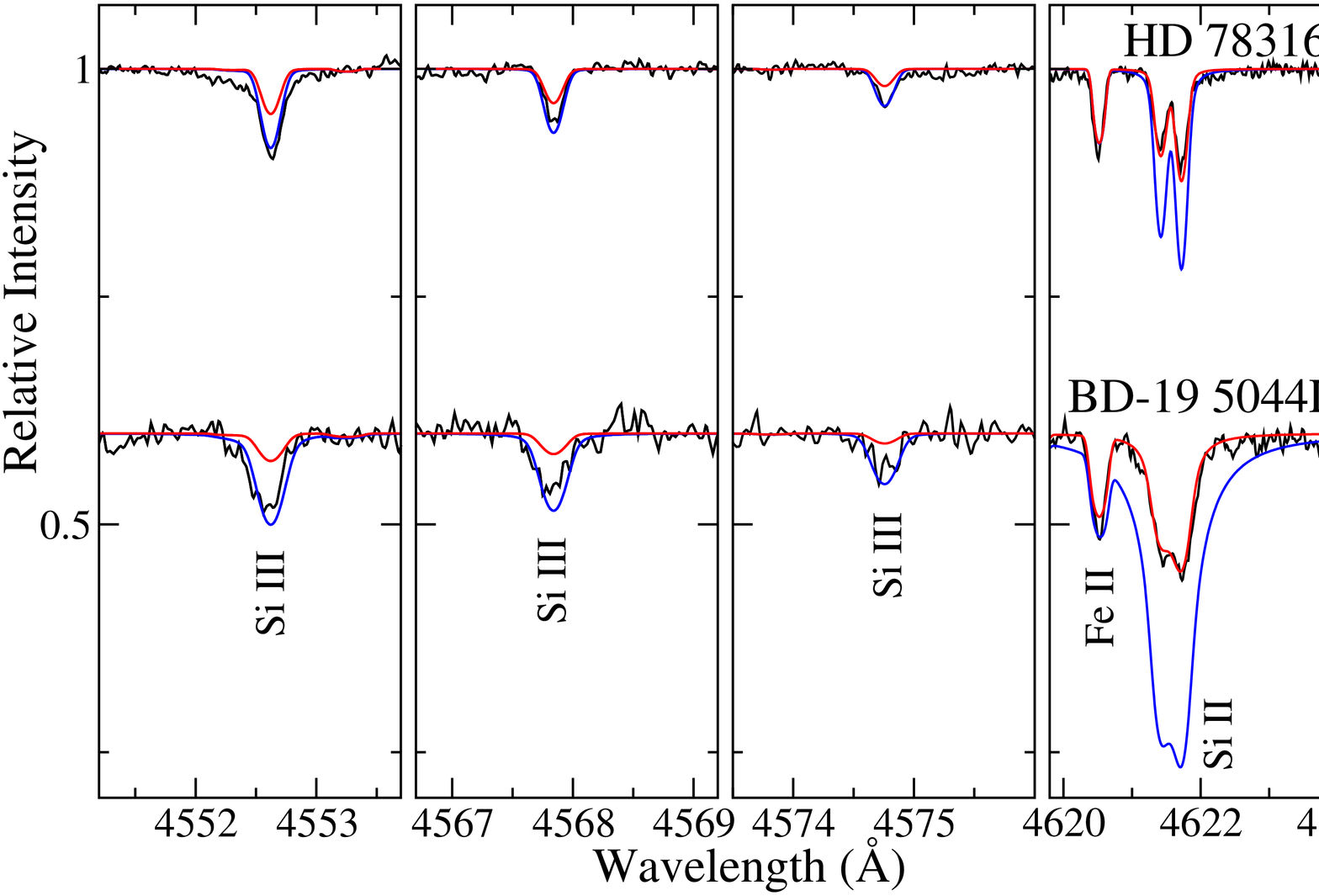}
 \caption{Model fits to lines of Si~{\sc ii} (right) and Si~{\sc iii} (left three panels) to the non-magnetic HgMn Star HD~78316 (upper spectrum) and the magnetic Bp star BD-19 5044L (lower spectrum).  The black lines are the observed spectra, red lines are the best fit model using the abundance derived from Si~{\sc ii} and the blue lines are the best fit model using the abundance derived from Si~{\sc iii}.} 
\label{si}
\end{center}
\end{figure}

\begin{table*}
\centering
\caption{Abundances derived for Ti, Cr, Fe, and the first and second ionization states of Si.  Abundance uncertainties are estimated to be about $\pm$0.2~dex for the magnetic Bp stars and $\pm$0.1~dex for the non-magnetic normal and HgMn stars.  Exceptional values are indicated in parentheses.  The second to last column indicates the difference between the Si~{\sc ii} and Si~{\sc iii} abundances and the last column denotes the significance $\Delta {\rm Si}/\sigma_{D}$ of the difference.}
\begin{tabular}{lrrrrrrrrrr}
\hline\hline
Star & \te\ (K) & \vsi\ (\kms) & $\xi$ (\kms) &  $\log$(Ti/H) & $\log$(Cr/H) & $\log$(Fe/H) & \multicolumn{2}{c}{$\log$(Si/H)} & $\Delta$Si & $\sigma_{D}$ \\
 & & & & & & & Si~{\sc ii} & Si~{\sc iii} &  & \\
\hline
\multicolumn{11}{c}{normal stars}\\
HD 222173 & 11800 & 67 $\pm$ 5 & $<$ 1.4 & $-7.80$ & $-6.64$ & $-4.96$ & $-4.59$(0.15) & $-4.03$(0.2) & 0.56 & 2.2\\
HD 22136 & 12700 & 15 $\pm$ 2 & 1.1 $\pm$ 0.3 & $-7.32$ & $-6.34$ & $-4.82$ & $-4.63$ & $-4.41$ & 0.22 & 1.4\\
HD 162586 & 12700 & 28 $\pm$ 2 & $<$ 1.6 & $-6.95$ & $-6.26$ & $-4.84$ & $-4.74$ & $-3.90$ & 0.84 & 5.9\\
HD 179761 & 12900 & 14 $\pm$ 2 & $<$ 1.5 & $-7.43$ & $-6.38$ & $-4.75$ & $-4.23$ & $-4.72$ & 0.49 & 3.5\\
HD 195810 & 13700 & 52 $\pm$ 4 & $<$ 1.1 & $-7.41$ & $-6.35$ & $-4.58$ & $-4.24$ & $-4.65$ & 0.41 & 2.9 \\
HD 170054 & 14500 & 25 $\pm$ 2 & $<$ 1.2 & $-7.46$ & $-6.67$ & $-4.95$ & $-4.72$ & $-4.21$ & 0.51 & 3.6\\
HD 160762 & 17500 & 6 $\pm$ 1 & 1 $\pm$ 1 &  --    &   --    & $-4.49$ & $-5.34$ & $-4.10$ & 1.24 & 8.9 \\
\hline
\multicolumn{11}{c}{HgMn stars}\\
HD 193452 & 10600 & 1.4 $\pm$ 1 & $<$ 1.0 & $-6.50$ & $-5.73$ & $-4.22$ & $-4.20$ & -- & -- & -- \\
HD 57608  & 10900 & 5.0 $\pm$ 1 & 1.3 $\pm$ 0.4 & $-8.08$ & $-6.81$ & $-5.31$ & $-4.78$ & $-4.82$ & $-0.04$ & 0.28\\
HD 27295  & 11800 & 4.9 $\pm$ 0.3 & 1.8 $\pm$ 0.5 & $-5.83$ & $-6.03$ & $-5.31$ & $-4.58$ & $-4.41$ & 0.17 & 1.2\\
HD 175640 & 12000 & 1.5 $\pm$ 1 & $<$ 1.7 & $-6.54$ & $-5.79$ & $-4.83$ & $-4.46$ & $-3.76$ & 0.70 & 5.0\\
HD 178065 & 12300 & 2.0 $\pm$ 1 & $<$ 1.5 & $-6.49$ & $-6.04$ & $-4.65$ & $-4.61$ & $-4.29$ & 0.32 & 2.3\\
HD 186122 & 12900 & 1.0 $\pm$ 0.5 & $<$ 1.4 & $-6.68$ & $-7.38$ & $-4.09$ & $-5.54$ & $-4.90$ & 0.64 & 4.5\\
HD 78316 & 13400 & 6.8 $\pm$ 0.5 & $<$ 1.6 & $-6.79$ & $-6.06$ & $-4.45$ & $-4.35$ & $-3.91$ & 0.44 & 3.2\\
\hline
\multicolumn{11}{c}{magnetic Bp stars}\\
HD 47116 & 11000 & 30 $\pm$ 2 & 0 & $-4.65$ & $-4.93$ & $-3.65$ & $-3.32$ & $-2.39$ & 0.93 & 3.3\\
HD 10840 & 11600 & 35 $\pm$ 5 & 0 & $-6.68$ & $-5.56$ & $-3.49$ & $-3.48$ & $-2.71$ & 0.77  & 2.7\\
HD 318107 & 11800 & 7 $\pm$ 1 & 0 & $-5.05$(0.1) & $-4.50$(0.15) & $-3.00$(0.2) & $-3.65$(0.15) & $-2.73$ & 0.92  & 3.7\\
HD 199728 & 12200 & 62 $\pm$ 6 & 0 & $-7.38$ & $-5.33$ & $-4.07$ & $-3.09$ & $-2.46$(0.25) & 0.63 & 2.0 \\
HD 223640 & 12300 & 31 $\pm$ 3 & 0 & $-6.89$ & $-4.65$ & $-3.42$ & $-3.42$ & $-2.17$ & 1.25 & 4.4\\
BD+00 1659 & 12500 & 7.0 $\pm$ 1 & 0 & $-6.69$ & $-5.02$ & $-3.79$ & $-3.53$ & $-2.51$ & 1.02 & 3.6\\
HD 304842 & 12500 & 65 $\pm$ 5 & 0 & $-6.82$ & $-6.16$ & $-4.34$ & $-3.65$ & $-2.91$(0.25) & 0.74 & 2.3 \\
HD 45583 & 12700 & 70 $\pm$ 6 & 0 & $-5.67$ & $-4.68$ & $-3.40$ & $-3.33$ & $-2.10$(0.25) & 1.23 & 3.8\\
BD-19 5044L & 12800 & 15 $\pm$ 3 & 0 & $-6.70$ & $-5.80$ & $-4.15$ & $-3.99$ & $-2.76$ & 1.23 & 4.3\\
BD+49 3789 & 12900 & 85 $\pm$ 5 & 0 & $-6.23$ & $-5.48$ & $-4.09$ & $-3.44$ & $-2.59$(0.3) & 0.85 & 2.4\\
HD 74168 & 12900 & 63 $\pm$ 4 & 0 & $-6.77$ & $-6.30$ & $-4.81$ & $-4.01$ & $-2.95$(0.25) & 1.06 & 3.3\\
HD 61045 & 13000 & 64 $\pm$ 3 & 0 & $-5.96$ & $-5.24$ & $-3.90$ & $-3.79$ & $-2.56$(0.25) & 1.23 & 3.8\\
HD 147010 & 13000 & 15 $\pm$ 2 & 0 & $-5.81$ & $-4.18$ & $-3.46$ & $-4.03$ & $-3.46$ & 0.57 & 2.0 \\
HD 133880 & 13000 & 103 $\pm$ 10 & 0 & $-4.86$ & $-4.49$ & $-3.30$ & $-2.68$ & $-1.23$(0.3) & 1.45 & 4.0\\
HD 137509 & 13100 & 20 $\pm$ 2 & 0 & $-4.62$ & $-4.20$ & $-3.19$ & $-3.52$ & $-2.02$ & 1.50 & 5.3\\
HD 74535 & 13600 & 45 $\pm$ 4 & 0 & $-6.25$ & $-5.60$ & $-3.98$ & $-4.25$ & $-2.53$ & 1.72 & 6.1\\
HD 49333 & 15100 & 68 $\pm$ 3 & 0 & $-5.97$ & $-5.66$ & $-4.18$ & $-3.99$ & $-2.50$(0.3) & 1.49 & 4.1\\
\hline
\label{silicon}
\end{tabular}
\end{table*}
 
In this Section we tabulate and discuss the results of our abundance analysis of the stars of the sample discussed above.

Table~\ref{silicon} presents the final mean abundances we find for each star. The first four columns list the star name (generally an HD Number),  and recall the \te, \vsi, and $\xi$ values from Table~\ref{parameters}. Next we tabulate abundance values found for Ti, Cr, and Fe. The abundances of these three elements had to be determined in order to evaluate and test the microturbulence parameter independently of the Si lines (for the non-magnetic stars only), to correctly include the effects of weak blends with the Si lines modelled, and to allow a precise determination of $v_{\rm R}$ and $v \sin i$.  For all stars, the abundance of iron was found from simultaneously fitting three lines of Fe~{\sc ii}: 4541, 4555, and 4583~\AA.  Similarly, where possible we derived the abundance of chromium from lines of Cr~{\sc ii} at 4558, 4588, and 4592~\AA\ and titanium from lines of Ti~{\sc ii} at 4565 and 4572~\AA.  The abundance of Fe and the value of $\xi$ for HD~160762 were taken from \citet{Nieva2012}, but the Si abundances were determined by us in the usual way. 

For the magnetic Bp stars, we estimate the uncertainties of the abundances of the Fe-peak elements to be $\pm$0.2~dex, a typical deviation from the derived abundance necessary to produce an unsatisfactory fit of the model to the observed spectra.  The uncertainty derived in the same manner for the non-magnetic stars suggests a smaller uncertainty of $\pm$0.1~dex.  Uncertainties that differ from the aforementioned values are noted in parentheses.  This is the case for stars with \vsi\ greater than about 60~\kms\ where line blending makes the derived abundances somewhat less certain. 

Following the Fe-peak elements, two columns list abundances derived from lines of Si~{\sc ii} and Si~{\sc iii} separately, for each star in our sample. The next column gives the difference of the two Si abundance determinations, in the sense Si~{\sc iii} -- Si~{\sc ii}, and the final column gives the difference divided by the estimated uncertainty of the difference, including only the fitting uncertainties discussed in this section. (Recall that the uncertainty in \te\ substantially increases the uncertainty in the abundance derived from Si~{\sc iii}, to at least $\pm 0.3$~dex.) 

It will immediately be seen that the values of $\log(N_{\rm Si}/N_{\rm H})$ are quite discordant as derived from the two different ionisation states of Si. The reader may wonder how robust these large differences are. Figure~\ref{si} illustrates the type of fits we obtain with magnetic and non-magnetic stars using abundances derived from lines of Si~{\sc ii} and Si~{\sc iii}. The situation for the lines of these two ions is mostly quite different than the fits found for the Fe-peak elements Ti, Cr and Fe, for which all lines of the elements are generally fit well with a single value of the abundance. In both the HgMn star (upper spectrum) and the magnetic  star (lower spectrum), the abundance that best fits the lines of multiplet (2) of Si~{\sc iii} at 4552-67-74~\AA\ (model fit in blue) predicts a Si~{\sc ii} line at 4621~\AA\ that is far too strong. Conversely, the abundance that fits the Si~{\sc ii} line at 4621~\AA, and the other Si~{\sc ii} lines modelled, leads to model lines for the Si~{\sc iii} lines that are much too weak. This discrepancy is really striking for the magnetic Bp star. It is present but at a less spectacular level for most of the HgMn stars, and even seems to be present for some of the normal stars. 

The sharp-lined HgMn stars, which have a roughly solar abundance of Si, allow us to establish the lower limit of \te\ for which Si~{\sc iii} is observable.  This was done by searching for the presence of $\lambda$4552, the strongest Si~{\sc iii} line in multiplet 2, which also includes $\lambda\lambda$4567, and 4574.  We are able to detect a very weak line of Si~{\sc iii} in HD~57608 (\te\ = 10900~K), however, in a slightly cooler star, HD~192452 (\te\ = 10600~K) we are unable to unambiguously detect any lines of Si~{\sc iii}. This threshold also depends somewhat on $\log g$, but for a solar abundance of Si, we can begin to detect Si~{\sc iii} in main sequence stars at $\te \approx 11\,000$~K. 
         
\section{Possible explanations of the Si~{\sc ii}-{\sc iii} anomaly}

The results presented in the previous section show that the abundances derived from the singly and doubly ionised states of Si are in general different. This difference is always of the order of one dex for the magnetic stars in our sample, and apparently varies from near zero to almost one dex for the normal and HgMn stars. The difference is always in the same sense: if the abundance is chosen to have a value consistent with the Si~{\sc ii} lines, the predicted strengths of the Si~{\sc iii} lines are always too small. 

It is first worth noticing that there are three stars for which no hint of discrepancy between the abundance of Si derived from the two ionization states. These are exactly the three stars for which we believe that we detect a non-zero microturbulence parameter $\xi$, and thus presumably the presence of atmospheric convection. It is not obvious what makes these stars different from other normal or HgMn stars of similar fundamental parameters, since we expect the presence or absence of atmospheric convective zones to vary smoothly with \te, $\log g$, and perhaps abundance. Thus our initial expectation was that at low \te\ values, where we expect weak convective mixing (but do not usually  securely detect it), all the stars might be reasonably mixed, or all not mixed. 

We also recall (Sec. 3.3) that the discrepancy between abundances deduced from Si~{\sc ii} and {\sc iii} decreases as \te\ increases. Thus a possible explanation of the discrepancy we find is a systematic underestimate of \te\ for almost all the stars of our sample. To explain the discrepancies found among normal and HgMn stars, the values of \te\ would have to be systematically underestimated by an amount of the order of 1000~K, which would reduce the discrepancy by about 0.5~dex. For the magnetic stars, the systematic error in \te\ needed would be of order 2000~K. Although such large errors are possible, they seem unlikely to us, and so we look for other explanations of the observed differences. 

We next consider two further possible explanations of the abundance discrepancy that we do detect in almost all the  stars in our sample: non-LTE effects, and Si abundance stratification. 

\subsection{Non-LTE effects in B-type stars}

\begin{table*}[ht]
\centering
\caption{Silicon abundances deduced for HD~160762 from lines of Si~{\sc ii} and {\sc iii} using the non-LTE equivalent width calculations of \citet{Becker1990}, and LTE models computed with {\sc zeeman}.}
\begin{tabular}{lrrrrr}
\hline\hline
 & & & \multicolumn{3}{c}{$\log$($N_{\rm Si}/N_{\rm H}$)}\\
 & & & \multicolumn{2}{c}{Becker \& Butler} & Zeeman \\
\cline{4-6}
Ion & Line (\AA) &  W$_{\lambda}$ (\AA) & $\xi = 0$~\kms & $\xi = 3$~\kms &  $\xi = 1$~\kms \\
\hline
Si~{\sc ii} & 4128 & 57.3 & $<$ $-5.0$ & $<$ $-5.0$ & $-5.13$\footnotemark[1] \\
            & 4130 & 63.2 & $<$ $-5.0$ & $<$ $-5.0$ & \\
            & 5041 & 40.1 & $<$ $-5.0$ & $<$ $-5.0$ & $-5.34$\footnotemark[2] \\
            & 5055 & 49.1 & $<$ $-5.0$ & $<$ $-5.0$ & \\
            & 5056 & 21.9 & $<$ $-5.0$ & $<$ $-5.0$ & \\
Si~{\sc iii} & 4552 & 70.1 & $-4.1$ & $-4.3$ & $-4.10$\footnotemark[3] \\
             & 4567 & 47.5 & $-4.3$ & $-4.5$ & \\
             & 4574 & 28.2 & $-4.2$ & $-4.5$ & \\
\hline
\label{beckerandbutler}
\end{tabular}
\end{table*}
\footnotetext[1]{Derived from $\lambda\lambda$4128 and 4130 simultaneously.}
\footnotetext[2]{Derived from $\lambda\lambda$5041, 5055, and 5056 simultaneously.}
\footnotetext[3]{Derived from $\lambda\lambda$4552, 4567, and 4574 simultaneously.}

Recent studies of non-LTE effects in the atmosphere of O and early B-type stars have been carried out \citep[e.g.][]{Przybilla2011, Nieva2012} testing the predictions of LTE and non-LTE line synthesis codes for main sequence stars with \te\ between about 15000 and 35000~K.  They show that a hybrid method, combining LTE model atmospheres with non-LTE line synthesis, is capable of yielding very accurate basic stellar parameters, and very satisfactory fit to most spectral lines. From comparisons of their non-LTE results to fully LTE synthesis, they conclude that pure LTE modelling yields meaningful results up to about a \te\ of 22000~K for a large number of spectral lines.  However, lines with this behaviour must be carefully identified.  \citet{Neiner2012} have been able to determine abundances of a number of elements for the star HD~96446 (\te\ = 21600 $\pm$ 800~K) by using only spectral lines that are approximately the right strength when computed with LTE synthesis and non-LTE abundances for standard stars such as $\alpha$~Pyx. 

Unfortunately, \citet{Neiner2012} find that lines of Si~{\sc ii} and {\sc iii} mostly are not among the lines that are predicted to have the observed strength in standard stars by LTE synthesis. Thus it appears that among early B type stars, non-LTE effects are probably important for Si. These results are consistent with comments by \citet{Przybilla2011}  who suggest that lines of silicon are strongly influenced by non-LTE effects above about 15000~K. 

We do not have the tools to carry out non-LTE synthesis for this project. However, we have included the star HD~160762 ($\te = 17500$~K) in our sample in order to explore the onset of non-LTE just above the temperature range of the stars we are studying. For this star, \citet{Nieva2012} find that the abundance of Si is essentially solar, with $\log N_{\rm Si}/N_{\rm H} = -4.49 \pm 0.05$. From our LTE analysis, it is found that the values of Si abundance as measured with lines of Si~{\sc ii} ($-5.34$) and Si~{\sc iii} ($-4.10$) are inconsistent with one another, and with the value found by \citet{Nieva2012}. The fact that with suitable choice of lines and appropriate non-LTE synthesis computations, the abundance of Si, like the other elements in this star that were measured by \citet{Nieva2012}, have solar abundances indicate that stratification is probably not significant in this star. Instead, the problems in this star with the lines we are using are probably non-LTE effects.  

We could try to test this conclusion by showing that non-LTE computation of our favoured lines in HD~160762 lead to an essentially solar, and consistent, abundance of Si. While we do not have the correct tools for this computation, we can try to use the non-LTE calculations of \citet{Becker1990}, who have computed non-LTE equivalent widths of a number of Si~{\sc ii} and {\sc iii} lines over a large grid of \te, $\log g$, and $\xi$. We have measured the equivalent widths of a number of spectral lines in HD~160762, and used the tables of \citet{Becker1990} to deduce corresponding Si abundances. The results of this exercise are shown in Table~\ref{beckerandbutler}, together with our own LTE abundances from the same lines. 

Assuming the correctness of the Si abundance of $-4.49$, we find that the Si abundance determined from lines of Si~{\sc iii}, using the computations of \citet{Becker1990} and interpolating to $\xi = 1$~\kms, is about $-4.25$, somewhat closer to the correct value than the LTE abundance of $-4.10$. In contrast, the non-LTE abundance deduced from lines of Si~{\sc ii} appears to be nearly one dex low, as is the LTE Si~{\sc ii} value. It thus seems that the computations of \citet{Becker1990} for Si~{\sc iii} multiplet (2) do not fully correct for non-LTE effects, and those for the optical lines of Si~{\sc ii}  do not even come close to correctly describing the non-LTE effects. (Note that the non-LTE treatment of Si~{\sc ii} is still apparently unsatisfactory today; the model non-LTE spectra for early B stars shown by \citet{Nieva2012} in their Figs 8 and 9 do not include any of the strong optical Si~{\sc ii} lines.)

Our conclusion is that by an effective temperature of about 17500~K non-LTE effects almost certainly do undermine Si abundance determinations from LTE spectrum synthesis, especially for lines of Si~{\sc ii}, but we are not at present able to quantify this effect, or even say how far down in effective temperature it reaches. The fact that \citet{Mashonkina2005,Mashonkina2009} find important non-LTE effects in formation of lines of Pr and Nd~{\sc ii} and {\sc iii} in magnetic Ap stars with \te\ as low as 7250~K suggests that non-LTE effects in Si could potentially be important well below the conventional limit of $\te \approx 15\,000$~K.

\subsection{Stratification}

A rather different possible explanation for the Si~{\sc ii} -- Si~{\sc iii} abundance discrepancies is the possibility that Si is not uniformly abundant with height in the atmosphere of the stars studied here. We have already mentioned this possibility earlier when we pointed out that the stars without the abundance discrepancy all show evidence of non-zero $\xi$ values, and hence are presumed to have convective mixing present in the atmosphere, suggesting that these stars at least may have atmospheres that are too well mixed to support any kind of vertical non-uniform abundances. However, for most of the stars in this survey, we have no strong evidence of vertical mixing, and in the case of the magnetic stars, the large energy density in the magnetic fields almost certainly quenches any convective turnover, so that vertical abundance stratification is quite possible. 

As discussed above, vertical stratification has already been detected in rare earths in cool magnetic Ap stars by \citet{Ryabchikova2001}, who found that Pr and Nd have substantially larger abundance relative to H in the upper atmospheres of pulsating magnetic (roAp) stars. This effect has been further studied by \citet{Kochukhov2009} and others cited in the Introduction. It now seems quite clear that stratification does occur to a very important extent in the cool magnetic Ap stars, and that it affects many chemical elements, including Si. Is there any evidence that stratification can explain the abundance discrepancies documented in this paper? 

We have carried out a series of simple experiments in which we introduce a simple smoothed step function of specified amplitude and at a specified optical depth $\tau_{\rm 5000}$ at 5000~\AA.  This was done using a slightly modified version of {\sc zeeman} that, given the optical depth and abundance change of the step, can optimise the fit to one or several spectral lines by adjusting the abundance below the step in the atmosphere. We mainly experimented on the two magnetic Bp stars BD+00 1659 and BD--19 5044L, both of which have \te\ values in the middle of our sample, around 12\,500~K, and such low values of \vsi\ that we can study the fit to line profiles, not merely to equivalent widths. This is a far simpler scheme than the detailed modelling described, for example, by \citet{Kochukhov2006b}. However, it enables us to explore simply the plausibility of stratification as an explanation of the Si~{\sc ii}--{\sc iii} discrepancy. 

We have explored mainly step functions in which the abundance decreases outwards. The few experiments with Si concentrated high in the atmosphere lead to even larger discrepancies between the abundances determined from the two ionisation states than with an assumed uniform abundance. We have roughly explored abundance decreases of 1 or 2~dex, located typically between $\tau_{\rm 5000} = 0.2$ and 0.8. It is found that most such models substantially reduce the discrepancy between the abundances deduced from the two ion states, often to below 0.5~dex. With a decrease of 1~dex in Si abundance at $\tau_{\rm 5000} \sim 0.4$ the discrepancy can be reduced to as low as 0.3~dex. Larger steps of 2~dex lead to nominal agreement between the two ionisation stages, but also lead to wings on the Si~{\sc ii} lines that are unacceptably broad (a problem that is only visible because we are studying stars of low \vsi). 

The conclusion of our limited experiments is that vertically non-uniform Si abundance, with a substantially lower abundance high in the atmosphere relative to $\tau_{\rm 5000} = 1$, can explain at least a large part of the discrepancy between abundances of Si deduced from the two ion states for magnetic Bp stars, and may entirely explain the Si~{\sc ii}--{\sc iii} differences for normal and HgMn stars. Clearly more detailed stratification modelling of individual stars would be very illuminating. Similarly, further study of the non-LTE behaviour of the Si ions in the \te\ range of 10\,000 to 15\,000~K would be very valuable.

\section{Discussion}

The results of our abundance analysis of the singly and doubly-ionised states of silicon in B-type stars are summarised as follows:\\
\begin{enumerate}
\item For a solar abundance of Si, we begin to detect Si~{\sc iii} in main sequence stars at \te\ $\sim$ 11000~K.
\item All magnetic Bp stars have a discrepancy between the abundances derived from lines of Si~{\sc ii} and Si~{\sc iii} with the latter being between 0.6 - 1.7~dex higher. The same discrepancy is observed in the non-magnetic stars, but to a much lesser extent, with an enhancement of the Si~{\sc iii} abundance by between 0.3 - 0.8~dex.
\item Only three non-magnetic stars do not exhibit a discrepancy between the abundance of Si derived from both ionisation states. These are exactly the stars where we detect a non-zero microturbulence parameter: HD~22136 (normal), HD~57608 (HgMn), HD~27295 (HgMn). This suggests the presence of convective mixing in the atmospheres of these stars, inhibiting the development of any stratified abundance distribution.
\item Possible errors in the atomic data (of the order of $\pm$0.1~dex) cannot explain this discrepancy. Although non-LTE effects in Si lines may be important, it is unclear to what extent these will influence the derived abundances below \te\ $\sim$ 15000~K.
\item Simple stratification models with a lower abundance high in the atmosphere relative to $\tau_{\rm 5000} = 1$ can explain a large part of the discrepancy between the derived abundances of Si~{\sc ii} and Si~{\sc iii} in the magnetic Bp stars, and may completely explain the phenomena in the normal and HgMn stars.
\end{enumerate} 

To fully understand the discrepancy noted between the abundances derived from the first and second ions of Si, more detailed stratification models are warranted. Further, studies of the non-LTE behaviour of Si below \te\ = 15000~K would reveal to what extent the observed discrepancy is influenced by non-LTE effects.

\begin{acknowledgements}
JDB and JDL are grateful for support by the Natural Sciences and Engineering Research Council of Canada.  
\end{acknowledgements}

\bibliographystyle{aa}
\bibliography{silicon}

\begin{thebibliography}{23}
\expandafter\ifx\csname natexlab\endcsname\relax\def\natexlab#1{#1}\fi

\bibitem[Babel(1992)]{Babel1992} Babel, J. 1992, A\&A, 258, 449

\bibitem[Bagnulo et al.(2006)]{Bagnulo2006} Bagnulo, S., Landstreet,
  J. D., Mason, E., et al. 2006, A\&A, 450, 777

\bibitem[{{Bailey} {et~al.}(2012){Bailey}, {Grunhut}, {Shultz}, {Wade},
  {Landstreet}, {Bohlender}, {Lim}, {Wong}, {Drake}, \& {Linsky}}]{Bailey2012}
{Bailey}, J.~D., {Grunhut}, J., {Shultz}, M., {et~al.} 2012, MNRAS, 2947

\bibitem[{{Bailey} {et~al.}(2011){Bailey}, {Landstreet}, {Bagnulo}, {Fossati},
  {Kochukhov}, {Paladini}, {Silvester}, \& {Wade}}]{Bailey2011}
{Bailey}, J.~D., {Landstreet}, J.~D., {Bagnulo}, S., {et~al.} 2011, A\&A, 535,
  A25

\bibitem[{{Becker} \& {Butler}(1990a)}]{Becker1990}
{Becker}, S.~R. \& {Butler}, K. 1990, A\&A, 235, 326



\bibitem[Kelleher \& Podobedova(2008)]{Kelleher2008} Kelleher, D. E. \&
  Podobedova, L. I. 2008, J. Phys. Chem. Ref. Data, 37, 1285

\bibitem[{{Kochukhov}(2006)}]{Koch2006}
{Kochukhov}, O. 2006, A\&A, 454, 321

\bibitem[Kochukhov et al.(2006)]{Kochukhov2006b} Kochukhov, O.,
  Tsymbal, V., Ryabchikova, T., Makaganyk, V., \& Bagnulo, S. 2006, A\&A,
  460, 831

\bibitem[Kochukhov et al.(2009)]{Kochukhov2009} Kochukhov, O.,
  Shulyak, D., \& Ryabchikova, T. 2009, A\&A, 499, 851

\bibitem[{{Kochukhov} \& {Bagnulo}(2006)}]{Kochukhov2006}
{Kochukhov}, O. \& {Bagnulo}, S. 2006, A\&A, 450, 763

\bibitem[{{Kunzli} {et~al.}(1997){Kunzli}, {North}, {Kurucz}, \&
  {Nicolet}}]{Kunzli1997}
{Kunzli}, M., {North}, P., {Kurucz}, R.~L., \& {Nicolet}, B. 1997, A\&AS, 122,
  51

\bibitem[{{Kupka} {et~al.}(1999){Kupka}, {Piskunov}, {Ryabchikova}, {Stempels},
  \& {Weiss}}]{vald4}
{Kupka}, F., {Piskunov}, N.~E., {Ryabchikova}, T.~A., {Stempels}, N.~C., \&
  {Weiss}, W.~W. 1999, A\&A, 138, 119

\bibitem[{{Kupka} {et~al.}(2000){Kupka}, {Ryabchikova}, {Piskunov}, {Stempels},
  \& {Weiss}}]{vald1}
{Kupka}, F.~G., {Ryabchikova}, T.~A., {Piskunov}, N.~E., {Stempels}, H.~C., \&
  {Weiss}, W.~W. 2000, Baltic Astronomy, 9, 590

\bibitem[{{Landstreet}(1988)}]{Landstreet1988}
{Landstreet}, J.~D. 1988, ApJ, 326, 967

\bibitem[{{Landstreet}(1998)}]{Landstreet1998}
{Landstreet}, J.~D. 1998, A\&A, 338, 1041

\bibitem[Landstreet(2011)]{Landstreet2011} Landstreet, J. D. 2011,
  A\&A, 528, A132

\bibitem[{{Landstreet} {et~al.}(2007){Landstreet}, {Bagnulo}, {Andretta},
  {Fossati}, {Mason}, {Silaj}, \& {Wade}}]{Landstreet2007}
{Landstreet}, J.~D., {Bagnulo}, S., {Andretta}, V., {et~al.} 2007, A\&A, 470,
  685

\bibitem[{{Landstreet} {et~al.}(1989){Landstreet}, {Barker}, {Bohlender}, \&
  {Jewison}}]{Landstreet1989}
{Landstreet}, J.~D., {Barker}, P.~K., {Bohlender}, D.~A., \& {Jewison}, M.~S.
  1989, ApJ, 344, 876

\bibitem[{{Landstreet} {et~al.}(2009){Landstreet}, {Kupka}, {Ford}, {Officer},
  {Sigut}, {Silaj}, {Strasser}, \& {Townshend}}]{Landstreet2009}
{Landstreet}, J.~D., {Kupka}, F., {Ford}, H.~A., {et~al.} 2009, A\&A,
503, 973

\bibitem[Landstreet et al.(2008)]{Landstreet2008} Landstreet, J. D.,
  Silaj, J., Andretta, V. et al. 2008, A\&A, 481, 465

\bibitem[Mashonkina et al.(2005)]{Mashonkina2005} Mashonkina, L.,
  Ryabchikova, T., \& Ryabtsev, A. 2005, A\&A, 441, 309

\bibitem[Mashonkina et al.(2009)]{Mashonkina2009}  Mashonkina, L.,
  Ryabchikova, T., Ryabtsev, A., \& Kildiyarova, R. 2009, A\&A. 495, 297

\bibitem[Napiwotzki et al.(1993)]{Napiwotzki1993} Napiwotzki, R.,
Sch\"{o}nberner, D., \& Wenske, V. 1993, A\&A, 268, 653

\bibitem[Neiner et al.(2012)]{Neiner2012} Neiner, C., Landstreet,
  J. D., Alecian, E. et al. 2012, A\&A, 546, 44

\bibitem[Netopil et al.(2008)]{Netopil2008} Netopil, M., Paunzen, E.,
  Maitzen, H. M., North, P., \& Hubrig, S. 2008, A\&A, 491, 545

\bibitem[{{Nieva} \& {Przybilla}(2012)}]{Nieva2012}
{Nieva}, M.-F. \& {Przybilla}, N. 2012, A\&A, 539, A143

\bibitem[Pandey et al.(2011)]{Pandey2011} Pandey, C. P., Shulyak,
  D. V., Ryabchikova, T., \& Kochukhov, O. 2011, MNRAS, 417, 444

\bibitem[{{Piskunov} {et~al.}(1995){Piskunov}, {Kupka}, {Ryabchikova}, {Weiss},
  \& {Jeffery}}]{vald3}
{Piskunov}, N.~E., {Kupka}, F., {Ryabchikova}, T.~A., {Weiss}, W.~W., \&
  {Jeffery}, C.~S. 1995, A\&A, 112, 525

\bibitem[{{Przybilla} {et~al.}(2011){Przybilla}, {Nieva}, \&
  {Butler}}]{Przybilla2011}
{Przybilla}, N., {Nieva}, M.-F., \& {Butler}, K. 2011, Journal of Physics
  Conference Series, 328, 012015

\bibitem[{{Ryabchikova} {et~al.}(1997){Ryabchikova}, {Piskunov}, {Kupka}, \&
  {Weiss}}]{vald2}
{Ryabchikova}, T.~A., {Piskunov}, N.~E., {Kupka}, F., \& {Weiss}, W.~W. 1997,
  Baltic Astronomy, 6, 244

\bibitem[{{Ryabchikova} {et~al.}(2001){Ryabchikova}, {Savanov}, {Malanushenko},
  \& {Kudryavtsev}}]{Ryabchikova2001}
{Ryabchikova}, T.~A., {Savanov}, I.~S., {Malanushenko}, V.~P., \&
  {Kudryavtsev}, D.~O. 2001, Astronomy Reports, 45, 382

\bibitem[{{Semenko} {et~al.}(2008){Semenko}, {Kudryavtsev}, {Ryabchikova}, \&
  {Romanyuk}}]{Semenko2008}
{Semenko}, E.~A., {Kudryavtsev}, D.~O., {Ryabchikova}, T.~A., \& {Romanyuk},
  I.~I. 2008, Astrophysical Bulletin, 63, 128

\bibitem[Shulyak et al.(2009)]{Shulyak2009} Shulyak, D., Ryabchikova,
  T., Mashonkina, L., \& Kochukhov, O. 2009, A\&A, 499, 879

\bibitem[Sigut \& Landstreet(1990)]{Sigut1990} Sigut, T. A. A. \&
  Landstreet, J. D. 1990, MNRAS, 247, 611

\bibitem[{{Wade} {et~al.}(2001{\natexlab{a}}){Wade}, {Bagnulo}, {Kochukhov},
  {Landstreet}, {Piskunov}, \& {Stift}}]{Wadeetal2001}
{Wade}, G.~A., {Bagnulo}, S., {Kochukhov}, O., {et~al.} 2001{\natexlab{a}},
  A\&A, 374, 265

\bibitem[{{Wade} {et~al.}(2001{\natexlab{b}}){Wade}, {Ryabchikova}, {Bagnulo},
  \& {Piskunov}}]{Wade2001}
{Wade}, G.~A., {Ryabchikova}, T.~A., {Bagnulo}, S., \& {Piskunov}, N.
  2001{\natexlab{b}}, in Astronomical Society of the Pacific Conference Series,
  Vol. 248, Magnetic Fields Across the Hertzsprung-Russell Diagram, ed.
  G.~{Mathys}, S.~K. {Solanki}, \& D.~T. {Wickramasinghe}, 373

\end{thebibliography}
\end{document}